\documentclass[11pt,citesort]{article}
\usepackage{amsfonts}
\usepackage{amsmath}
\usepackage{amssymb}
\usepackage{latexsym}
\usepackage{color}
\usepackage{colordvi}
\usepackage{epsfig}
\usepackage{array}
\usepackage{pifont}
\usepackage{cite}
\usepackage{amsxtra}
\usepackage{axodraw}
\usepackage{geometry}
\def\krto{ {\,\,\lower .8ex\hbox {$\longrightarrow \atop k \rightarrow 0$}\,\,}}
\def\Section#1{\section{#1}\hspace{\parindent}}
\def\subSection#1{\subsection{#1}\hspace{\parindent}}

\def\bea{\begin{eqnarray} }
\def\beq{\begin{eqnarray} }

\def\eea{\end{eqnarray}}
\def\eeq{\end{eqnarray}}
\def\eq#1{eq.~(\ref{#1})}

\newcommand{\ghostSD}{\begin{picture}(150,25)(0,0)
\SetWidth{1.2}
\DashArrowLine(12.5,0)(37.5,0){5}
\DashArrowLine(37.5,0)(75,0){5}
\DashLine(75,0)(112.5,0){5}
\DashArrowLine(112.5,0)(137.5,0){5}
\SetWidth{1}
\Vertex(112.5,0){2}
\GlueArc(75,0)(37.5,0,90){-4}{6}
\GlueArc(75,0)(37.5,90,180){-4}{6}
\CCirc(75,0){5}{Black}{Yellow}
\CCirc(75,37.5){5}{Black}{Yellow}
\CCirc(37.5,0){5}{Black}{Yellow}
\Text(20,-10)[l]{a,k}
\Text(50,15)[l]{d,$\nu$}
\Text(100,-10)[l]{e}
\Text(100,15)[r]{f,$\mu$}
\Text(50,-10)[l]{c,q}
\Text(120,-10)[l]{b,k}
\Text(75,48)[c]{q-k}
\end{picture}}
\newcommand{\ghostDr}{\begin{picture}(100,25)(0,0)
\SetWidth{1.2}
\DashArrowLine(12.5,0)(50,0){5}
\DashArrowLine(50,0)(87.5,0){5}
\CCirc(50,0){5}{Black}{Yellow}
\Text(12.5,-10)[l]{a}
\Text(87.5,-10)[r]{b}
\Text(50,-10)[c]{k}
\end{picture}}
\newcommand{\ghostBr}{\begin{picture}(100,25)(0,0)
\SetWidth{1.2}
\DashArrowLine(12.5,0)(87.5,0){5}
\Text(12.5,-10)[l]{a}
\Text(87.5,-10)[r]{b}
\Text(50,-10)[c]{k}
\end{picture}}
\newcommand{\gluonSDi}{\begin{picture}(112.5,18.75)(0,0)
\SetScale{0.75}
\SetWidth{1.2}
\Gluon(12.5,0)(37.5,0){-4}{2}
\Gluon(37.5,0)(75,0){-4}{3}
\Gluon(75,0)(112.5,0){-4}{3}
\Gluon(112.5,0)(137.5,0){-4}{2}
\SetWidth{1}
\Vertex(112.5,0){2}
\GlueArc(75,0)(37.5,0,90){-4}{6}
\GlueArc(75,0)(37.5,90,180){-4}{6}
\CCirc(75,0){5}{Black}{Yellow}
\CCirc(75,37.5){5}{Black}{Yellow}
\CCirc(37.5,0){5}{Black}{Yellow}
\end{picture}}

\newcommand{\gluonSDii}{\begin{picture}(112.5,18.75)(0,0)
\SetScale{0.75}
\SetWidth{1.2}
\Gluon(15,-5)(75,-5){-3}{4}
\Gluon(75,-5)(135,-5){-3}{4}
\SetWidth{1}
\GlueArc(75,18.75)(18.75,-90,90){3}{4}
\GlueArc(75,18.75)(18.75,90,270){3}{4}
\CCirc(75,-2.5){5}{Black}{Yellow}
\CCirc(75,37.5){5}{Black}{Yellow}
\end{picture}}

\newcommand{\gluonSDiib}{\begin{picture}(112.5,18.75)(0,0)
\SetScale{0.75}
\SetWidth{1.2}
\Gluon(15,-5)(75,-5){-3}{4}
\Gluon(75,-5)(135,-5){-3}{4}
\SetWidth{1}
\GlueArc(75,18.75)(18.75,-90,90){3}{4}
\GlueArc(75,18.75)(18.75,90,270){3}{4}
\Vertex(75,-2.5){2}
\CCirc(75,37.5){5}{Black}{Yellow}
\end{picture}}

\newcommand{\gluonSDiii}{\begin{picture}(112.5,18.75)(0,0)
\SetScale{0.75}
\SetWidth{1.2}
\Gluon(12.5,0)(37.5,0){-4}{2}
\DashLine(37.5,0)(75,0){4}
\DashLine(75,0)(112.5,0){4}
\Gluon(112.5,0)(137.5,0){-4}{2}
\SetWidth{1}
\Vertex(112.5,0){2}
\DashCArc(75,0)(37.5,0,90){4}
\DashCArc(75,0)(37.5,90,180){4}
\CCirc(75,3){5}{Black}{Yellow}
\CCirc(75,37.5){5}{Black}{Yellow}
\CCirc(37.5,3){5}{Black}{Yellow}
\end{picture}}

\newcommand{\gluonSDiv}{\begin{picture}(112.5,18.75)(0,0)
\SetScale{0.75}
\SetWidth{1.2}
\Gluon(15,-5)(75,-5){-3}{4}
\Gluon(75,-5)(135,-5){-3}{4}
\SetWidth{1}
\DashCArc(75,18.75)(18.75,-90,90){4}
\DashCArc(75,18.75)(18.75,90,270){4}
\CCirc(75,-2.5){5}{Black}{Yellow}
\CCirc(75,37.5){5}{Black}{Yellow}
\end{picture}}

\newcommand{\gluonSDv}{\begin{picture}(90,18.75)(0,0)
\SetScale{0.6}
\SetWidth{1.2}
\Gluon(12.5,0)(37.5,0){-4}{2}
\Gluon(37.5,0)(75,0){-4}{3}
\Gluon(75,0)(112.5,0){-4}{3}
\Gluon(112.5,0)(137.5,0){-4}{2}
\SetWidth{1}
\Vertex(112.5,0){2}
\GlueArc(75,0)(37.5,0,90){-4}{6}
\GlueArc(75,0)(37.5,90,180){-4}{6}
\GlueArc(75,0)(37.5,180,270){-4}{6}
\GlueArc(75,0)(37.5,270,360){-4}{6}
\CCirc(75,0){5}{Black}{Yellow}
\CCirc(75,37.5){5}{Black}{Yellow}
\CCirc(75,-37.5){5}{Black}{Yellow}
\CCirc(37.5,0){5}{Black}{Yellow}
\end{picture}}

\newcommand{\gluonSDvi}{\begin{picture}(112.5,18.75)(0,0)
\SetScale{0.6}
\SetWidth{1.2}
\Gluon(12.5,0)(37.5,0){-4}{2}
\Gluon(112.5,0)(137.5,0){-4}{2}
\Gluon(48.48,26.52)(112.5,0){-4}{5}
\SetWidth{1}
\Vertex(112.5,0){2}
\GlueArc(75,0)(37.5,0,90){-4}{6}
\GlueArc(75,0)(37.5,90,180){-4}{6}
\GlueArc(75,0)(37.5,180,270){-4}{6}
\GlueArc(75,0)(37.5,270,360){-4}{6}
\CCirc(80.49,13.26){5}{Black}{Yellow}
\CCirc(48.48,26.52){5}{Black}{Yellow}
\CCirc(75,37.5){5}{Black}{Yellow}
\CCirc(75,-37.5){5}{Black}{Yellow}
\CCirc(37.5,0){5}{Black}{Yellow}
\end{picture}}

\title{On the massive gluon propagator, the PT-BFM scheme and the low-momentum behaviour of decoupling 
and scaling DSE solutions}

\author{J. Rodr\'{\i}guez-Quintero}
\date{}

\begin{document} %\date{\today}

\maketitle

\begin{center}
Dpto. F\'isica Aplicada, Fac. Ciencias Experimentales,\\
Universidad de Huelva, 21071 Huelva, Spain
\end{center}

\begin{abstract}

We study the low-momentum behaviour of Yang-Mills propagators obtained from 
Landau-gauge Dyson-Schwinger equations (DSE) in the PT-BFM scheme. We compare 
the ghost propagator numerical results with the analytical ones obtained  
by analyzing the low-momentum behaviour of the ghost propagator DSE in Landau gauge, 
assuming for the truncation a constant ghost-gluon vertex
% as it is extensively done, 
and a simple model for a massive gluon propagator.
The asymptotic expression obtained for the regular or decoupling  ghost dressing function 
up to the order ${\cal O}(q^2)$ is proven to fit pretty well the numerical PT-BFM 
results. Furthermore, when the size of the coupling renormalized at 
some scale approaches some critical value, the numerical PT-BFM propagators tend 
to behave as the scaling ones. We also show that the scaling solution, 
implying a diverging ghost dressing function, cannot be a DSE solution 
in the PT-BFM scheme but an unattainable limiting case. 

\end{abstract}

%\pacs{12.38.Aw, 12.38.Lg, 12.38Gc }

\begin{flushright}
{\small UHU-FP/10-025}\\
%%{\small CPHT RR 038.0605}\\
%{\small LPT-Orsay/10-nn}\\
\end{flushright}

%\vfill
%\newpage

%%%%%%%%%%%%%%%%%%%%%%%%%%%%%%%%%%%%%%%%%%%%%%%%%%%%%%%%%
%% end of title page
%%%%%%%%%%%%%%%%%%%%%%%%%%%%%%%%%%%%%%%%%%%%%%%%%%%%%%%%%

%%%%%%%%%%%%%%%%%%%%%%%%%%%%%%%%%%%%%%%%%%%%%%%%%%%%%%%%%
%% body of the paper
%%%%%%%%%%%%%%%%%%%%%%%%%%%%%%%%%%%%%%%%%%%%%%%%%%%%%%%%%

\Section{Introduction}
%\alinea

The low-momentum behaviour of the Yang-Mills propagators derived either from 
the tower of Dyson-Schwinger equations (DSE) or from Lattice simulations in 
Landau gauge has been a very interesting and hot topic for the 
last few years. It seems by now well established that, if we assume 
in the vanishing momentum limit a ghost dressing function behaving as 
$F(q^2) \sim (q^2)^{\alpha_F}$ and a gluon propagator as 
$\Delta(q^2) \sim (q^2)^{\alpha_G-1}$ (or, by following a notation commonly used,
a gluon dressing function as $G(q^2)= q^2 \Delta(q^2) \sim (q^2)^{\alpha_G}$), 
two classes of solutions may emerge (see, for instance, the discussion 
of refs.~\cite{Boucaud:2008ji,Boucaud:2008ky}) from the DSE:
(i) those, dubbed {\it ``decoupling''}, where $\alpha_F=0$ and the suppression of 
the ghost contribution to the gluon propagator DSE results in a massive gluon 
propagator (see \cite{Aguilar:2006gr,Aguilar:2008xm} and references therein); 
and (ii) those, dubbed {\it ``scaling''}, where $\alpha_F \neq 0$ 
and the low-momentum behaviour of both gluon and ghost propagators 
are related by the coupled system of DSE through the condition $2 \alpha_F+\alpha_G = 0$ 
implying that $F^2(q^2)G(q^2)$ goes to a non-vanishing constant when $q^2 \to 0$ 
(see \cite{Alkofer:2000wg,Fischer:2008uz} and references therein).
As a matter of fact, $F^2(q^2)G(q^2)$, which gives the perturbative running for the
coupling constant renormalized in Taylor-scheme~\cite{Boucaud:2008gn}, is a good 
quantity in discriminating the kind of solutions we deal with. It is worth to 
remember that, despite the widely accepted nomeclature for the classes 
of solutions, neither a scale invariance nor a decoupling of the IR dynamics for the theory
can be inferred from the low-momentum behaviour of such a Taylor coupling.

How both types of IR solutions for Landau gauge DSE emerge and how the transition between them occurs, 
being governed by the size of the coupling taken as an integration boundary condition at 
the renormalization momentum, was initially discussed in ref.~\cite{Boucaud:2008ji} through the analysis of a
ghost propagator DSE combined with a gluon propagator taken from lattice computations. 
It should be remembered that one needs to know the QCD mass scale to predict the QCD coupling at any momentum. 
This mass scale should be of course supplied to get a particular solution from DSE and can be 
univocally related to the boundary condition needed, after applying a trunctation scheme, to solve 
the equation.  
The existence of a {\it critical} value for the coupling at any renormalization momentum was suggested 
by that partial analysis. No solution was proved to exist for any coupling bigger than the critical one and 
the unique scaling solution\footnote{The authors of \cite{Fischer:2006vf} proved there, once the scaling behaviour 
is assumed, the uniqueness for Yang-Mills infrared solutions} 
appeared to emerge when the coupling took that critical value.
Later, the authors of ref.~\cite{Fischer:2008uz} confirmed, by the analysis of the tower of DSE truncated 
within two different schemes and also in the framework of the functional renormalization group, that 
the boundary condition for the DSE integration determined whether a decoupling or the scaling solution 
occurs. A similar analysis have been recently done in the Coulomb gauge~\cite{Watson:2010cn} leading to 
the same pattern of ref.~\cite{Boucaud:2008ji}, although the authors interpreted the boundary condition in terms 
of the gauge-fixing ambiguity (see also \cite{Epple:2007ut}). 
Furthermore, an analytic study based on the pinch technique in ref.~\cite{Cornwall:2009ud} shows that, within some 
approximations, there is a lower limit for the gluon mass, which the IR singularities of QCD still persist below, 
that can be also interpreted as an upper limit to the coupling. 
Very recently also, a next-to-leading low-momentum asymptotic formula 
for the decoupling ghost dressing function solutions was obtained by studying the ghost propagator DSE with the 
assumption, for the truncation, of a constant ghost-gluon vertex and of a simple model 
for a massive gluon propagator~\cite{Boucaud:2010gr}. In this asymptotic formula, the 
ghost-propagator low-momentum behaviour appeared to be regulated by 
the zero-momentum effective charge in Taylor scheme~\cite{Aguilar:2009nf} 
and by the Landau-gauge gluon mass scale.

In the present note, the work of ref.~\cite{Boucaud:2008ji} will be extended by the analysis of the 
results~\cite{Aguilar} obtained by solving the coupled system of Landau gauge ghost and gluon propagators DSE 
within the framework of the pinching technique in the background field method~\cite{Binosi:2002ft} (PT-BFM). 
Our main goal is investigating whether the same pattern for regular (decoupling) and critical (scaling) solutions in 
ref.~\cite{Boucaud:2008ji} is also found for the PT-BFM solutions. In the PT-BFM scheme, 
the zero-momentum ghost dressing function will be seen to diverge when the coupling approaches some critical value, as it 
should be expected for the scaling solution ($\alpha_F \neq 0$). This seems to support the 
suggestion of a transition from one to another solutions controled by 
the size of the coupling approaching a critical value~\cite{Boucaud:2008ji}. 
The authors of ref.~\cite{Fischer:2008uz} obtained similar results but they applied the zero-momentum 
ghost propagator as the boundary condition for the 
DSEs integration and missed its connection with the value of the coupling at the renormalization momentum 
({\it i.e.} the particular value of $\Lambda_{\rm QCD}$ one applies to build the solutions) or the critical coupling 
the scaling behaviour requires to emerge\footnote{These authors furthermore invoked the renormalization 
group invariance to claim that such a critical coupling does not appear. Nevertheless, the renormalization group 
invariance only requires for the critical values of the coupling at any two fixed renormalized momenta  
to be connected by the appropriate renormalization group running (the same, of course, for any decoupling 
solution). Indeed, the renormalization flow for the coupling can be defined by $\alpha(q^2) = \alpha(\mu^2) F^2(q^2) G(q^2)$ 
to satisfy this and it is found to agree, in the perturbative domain, with the perturbative running given by the $\beta$-function 
in Taylor scheme~\cite{Boucaud:2008gn}.}. This connection is an important ingredient because it provides us with a manner, 
through a comparison with the physical strong coupling, to discuss whether the scaling critical DSE solution could 
be allowed by the data. We will also compare the low-momentum analytic results of ref.~\cite{Boucaud:2010gr} with 
the PT-BFM results.  In particular, the next-to-leading asymptotic 
formula for the ghost dressing function  with $\alpha_F=0$ will be shown to nicely 
describe the low-momentum PT-BFM results for different values of the coupling 
taken at the renormalization point, as a boundary condition for the DSE integration. 
It should be emphasized that the ansatz for the massive gluon propagator applied to derive the 
analytical results in ref.~\cite{Boucaud:2010gr} has nothing to do either with the numerical 
analysis based on lattice results in ~\cite{Boucaud:2008ji} or with the one based on PT-BFM results in this paper, 
both leading to obtain a critical coupling. These analytical low momentum results are applied in 
this paper only for comparative purposes. However, as a result of the comparison, this boundary condition will be put in 
direct relation to the zero-momentum effective charge in Taylor scheme~\cite{Aguilar:2009nf} 
and to the Landau gauge gluon mass. Finally, we will also argue that a scaling solution cannot exist as a solution of the 
coupled system of PT-BFM DSE, although the PT-BFM solutions tend to it when the coupling aproaches the critical value. 
More generally, a diverging ghost dressing function cannot be obtained when the gluon propagator is massive ($\alpha_G=1$).
Of course, this last claim is not a new result: it is well-known that the scaling solution only can emerge 
if the ghost dominance in the gluon propagator DSE, after assuming $\alpha_F < 0 $, implies the gluon 
propagator to vanish at zero momentum~\cite{Alkofer:2000wg,Fischer:2008uz}. However, it is not worthless to
emphasize that, for the gluon propagator, being massive implies not to observe 
the relation $2 \alpha_F + \alpha_G=0$ and is therefore a sufficient condition for the scaling 
solution not to appear. It should be also pointed out that LQCD results 
(see~\cite{Cucchieri:2007md,Bogolubsky:2007ud,IlgenGrib,Boucaud:2005ce,Oliveira:2010xc,Bornyakov:2009ug} 
and references therein), pinching technique results (see, for instance, 
\cite{Cornwall,Binosi:2002ft,Sauli:2009se,Cornwall:2009ud}), 
refined Gribov-Zwanziger~\footnote{In addition, K-I. Kondo triggered very recently an interesting 
discussion about the Gribov horizon condition and its implications on the Landau-gauge 
Yang-Mills infrared solutions~\cite{Kondo:2009ug,Dudal:2009xh,Aguilar:2009pp,Boucaud:2009sd}.} 
formalism (see~\cite{Dudal:2007cw}) or 
other approaches like the infrared mapping of $\lambda \phi^4$ and Yang-Mills 
theories in ref.~\cite{Frasca:2007uz} or the massive extension of the Fadeev-Popov 
action in ref.~\cite{Tissier:2010ts} appear to support a massive gluon propagator.

We organized this note as follows: first, we briefly review the low-momentum behaviour for the propagator solutions 
of the ghost propagator DSE in section \ref{twosol}; we then compare the PT-BFM results with the low-momentum analytical 
expression and discuss their dependence with the size of the coupling at the renormalization point, taken as a 
boundary condition for DSE integration, in section \ref{comparing}; and we finally conclude 
in section \ref{conclu}.

\Section{The two kinds of solutions of the ghost propagator Dyson-Schwinger equation}\label{revisiting}
\label{twosol}
%\alinea 

As was explained in detail in refs.~\cite{Boucaud:2008ky,Boucaud:2010gr}, the low-momentum behavior for 
the solutions of the Dyson-Schwinger equation for the ghost
propagator (GPDSE), which can be written diagrammatically as

\vspace{\baselineskip}
\begin{small}
\bea
\left(\ghostDr\right)^{-1}% 
=
\left(\ghostBr\right)^{-1}%
- 
\ghostSD \ , 
%\nonumber
\label{ghostSD}
\eea\end{small}%

\noindent can be obtained in Landau gauge by procceeding as follows: 

we consider eq.~(\ref{ghostSD}) for two different (although parallel) 
external ghost momenta, $k$ and $p$, such that
$p^2-k^2=\delta^2 k^2$ ($\delta$ being an extra parameter that, 
for the sake of simplicity, will be taken to be small enough as to 
expand on it around 0) and subtract them, as a regularization prescription 
for not to have to deal with any UV cut-off. Then, the ghost dressing function, 
$F(k^2)$, and the gluon propagator form factor, $\Delta(k^2)$, are renormalized by 
applying the MOM prescription, 

\beq
F_R(\mu^2) \ = \ \mu^2 \Delta_R(\mu^2) \ = \ 1 \ ,
\eeq

\noindent 
where $\mu^2$ is the subtraction point; and, as explained in \cite{Boucaud:2008ky,Boucaud:2010gr}, 
we choose for the ghost-gluon vertex, 

\beq
\widetilde{\Gamma}_\nu^{abc}(-q,k;q-k) \ = \
i g_0 f^{abc} \left( \ q_\nu H_1(q,k) + (q-k)_\nu H_2(q,k) \ \right) \ ,
\label{DefH12}
\eeq

\noindent
to apply the MOM prescription in Taylor kinematics 
({\it i.e.} with a vanishing incoming ghost momentum) 
and assume the non-renormalizable bare ghost-gluon form factor, $H_1(q,k)=H_1$, 
to be constant (at least in the low-momentum regime for the incoming ghost). 
Thus, after being cast into a renormalized form and the above-mentioned 
subtraction, the GPDSE reads

\beq
\frac{1}{F_R(k^2)} - \frac{1}{F_R(p^2)}  
\ = \  
N_C \ g_R^2(\mu^2) \ H_1 \ I(k^2) \ ,
\label{SDRS}
\eeq

\noindent
where

\beq
\label{LamInf}
I(k^2) \ = \  \ \int \frac{d^4 q}{(2\pi)^4} 
\left( \rule[0cm]{0cm}{0.8cm}
\frac{F_R(q^2)}{q^2} \left(\frac{(k\cdot q)^2}{k^2}-q^2\right) %\right. 
%\nonumber \\ 
% &\times& 
%\left. 
\left[ \rule[0cm]{0cm}{0.6cm}
\frac{\Delta_R\left((q-k)^2\right)}{(q-k)^2} -  
\frac{\Delta_R\left((q-p)^2\right)}{(q-p)^2} 
\rule[0cm]{0cm}{0.6cm} \right]
\rule[0cm]{0cm}{0.8cm} \right) \ 
\eeq

\noindent 
and where the ghost dressing function and gluon propagator should be understood as renormalized at 
the subtraction momentum, $\mu^2$, and $g_R(\mu^2)$ is the gauge coupling renormalized 
at a subtraction point with Taylor kinematics (Taylor MOM scheme).
Then, we cut integral domain of $I(k^2)$ in eq.~(\ref{LamInf}) into two pieces 
by introducing some new momentum scale, $q_0^2$, below which
both ghost and gluon are assumed to be well described by the following 
ansatze

\beq\label{gluonprop}
\Delta_R(q^2) &=& \frac{B(\mu^2)}{q^2 + M^2} \ 
\simeq \frac{B(\mu^2)}{M^2} \left( 1 - \frac{q^2}{M^2} + \cdots \right) \ ,
\\ \label{dress}
F_R(q^2) &=& A(\mu^2) \left( \frac{q^2}{M^2} \right)^{\alpha_F} \left( 1 + \cdots 
\rule[0cm]{0cm}{0.6cm} \right) \ .
\eeq

\noindent 
Thus, we shall look for the ghost dressing function, $F_{\rm IR}$,
its leading behaviour being parameterized through a general power law behaviour 
where $\alpha_F > -2$ to keep the integral $I(k^2)$ infrared convergent, and 
for a massive~\footnote{This is the massive gluon 
propagator where the gluon running mass~\cite{Lavelle:1991}, $M(q^2)$, appears to be approximated by 
its frozen value at vanishing momentum, $M(0)$. It should be also noted that, provided 
that the gluon propagator is to be multiplicatively renormalized, the mass scale, $M=M(0)$, 
does not depend on renormalization scale, $\mu^2$.} gluon propagator, that implies of course a 
power law with $\alpha_G=1$, as the current lattice data seems to point to. 
It is well-known that the low-momentum behaviour of the 
integral $I(k^2)$ in \eq{LamInf} is dominated by the result of the integration over the IR domain, 
over $q^2 < q_{0}^2$, and it can be expanded on $\delta^2 = p^2/k^2-1$ around zero with 
the subtraction momentum $\mu^2$ kept fixed, as explained in 
ref.~\cite{Boucaud:2008ky,Boucaud:2010gr}, to give
\beq
I(k^2)  \ \simeq \  I_{\rm IR}(k^2) 
&\simeq &
- \frac{\delta^2}{M^{2+2\alpha_F}} \frac{2 A(\mu^2) B(\mu^2)}{(2\pi)^3}  
\nonumber \\
&\times& \sum_{i=0}^\infty \ (4 k^2)^i  C_i \ \displaystyle \int_0^{q_{0}} q^{3+2i+2\alpha_F} dq \
K_i(q^2;k^2,M^2) \ + \ {\cal O}(\delta^4)  
\label{IRKi}
\eeq
where
\beq\label{Rser}
K_i(q^2;k^2,M^2) &=&   \frac {i} {(q^2+ k^2 + M^2)^{2i+1}} - \frac {i} 
{(q^2+ k^2)^{2i+1}}  \nonumber \\ 
&&  \ + \ \ k^2 \left( \frac {2i+1} {(q^2+ k^2)^{2i+2}} 
- \frac {2i+1} {(q^2+ k^2 + M^2)^{2i+2}} \right) 
\eeq
and
\beq
C_i \ = \ \frac{12 \pi^2 4^{i}}{\Gamma(-3/2-i) \Gamma(1/2-i) \Gamma(5+2i)} \ .
\eeq

Now, the two possible cases for the low-momentum behaviour of the 
GPDSE solutions will be separately analyzed in the following by applying \eq{IRKi} 
to \eq{SDRS}.

\subSection{Critical case: scaling solution}

When $\alpha_F < 0$, one can perform the change $t=q^2/k^2$ and integrate over $t$, the 
integration in \eq{IRKi} for any term of \eq{Rser} being (one by one) convergent 
as the limit $q_0^2/k^2 \to \infty$ is considered for the upper bound. 
Thus, the small-momentum asymptotical behaviour of the r.h.s of \eq{SDRS} 
is given by  
\beq\label{IRl0}
I_{\rm IR}(k^2) 
&\simeq& 
- \delta^2 \ \left(\frac{k^2}{M^2}\right)^{1+\alpha_F} \frac{2 A(\mu^2) B(\mu^2)}{(2\pi)^3} 
 \sum_{i=0}^\infty \ 4^i  \frac{C_i} 2 \ 
\nonumber \\
&\times&
\left( - i \ \int_0^{\infty}dt \frac{t^{1+i+\alpha_F}}{(1+t)^{2i+1}} 
\ + (2i+1) \  
\int_0^{\infty}dt \frac{t^{1+i+\alpha_F}}{(1+t)^{2i+2}} 
\right) \nonumber \\
&+& {\cal O}(\delta^4) \ ,
\eeq
while its l.h.s. behaves as:
\beq
\frac 1 {F_R(k^2)} - \frac 1 {F_R(p^2)} \ \simeq \ 
\delta^2 \frac{\alpha_F}{A(\mu)} \left( \frac{k^2}{M^2} \right)^{-\alpha_F}
\ + \ {\cal O}(\delta^4)
\eeq
Then, we should conclude both
\begin{itemize}
\item[(i)] that $-\alpha_F = 1+\alpha_F \ \ \Rightarrow \ \ \alpha_F=-1/2 $,
\item[(ii)] and, given that 
\beq
\sum_{i=0}^\infty \ 4^i \ C_i \ 
\left( - i \ \int_0^{\infty}dt \frac{t^{\frac 1 2 +i}}{(1+t)^{2i+1}} 
\ + (2i+1) \  
\int_0^{\infty}dt \frac{t^{\frac 1 2 +i}}{(1+t)^{2i+2}} 
\right) \ = \ \frac {2 \pi} 5 \ ,
\label{rela1}
\eeq
one obtains
\beq\label{crit-be}
N_C  g_R^2(\mu^2) H_1
A^2(\mu^2) B(\mu^2)
\simeq
10 \pi^2 \ .
\eeq
\end{itemize}

Thus, we know the asymptotic behaviour for the ghost dressing function in this case 
to be :
\beq 
F_R(q^2) \simeq 
\frac {\pi}{g_R(\mu^2)} 
\left( 
\frac{10 }{N_C H_1 \Delta_R(0)} 
\right)^{1/2}
\ \left(\frac 1 {q^2} \right)^{1/2}
\eeq
This is a so-called scaling solution where, in particular, the low-momentum behavior of 
the massive gluon propagator forces the ghost dressing function to diverge at low-momentum 
through the requirement~\footnote{This is of course a particular case, with $\alpha_G=1$,
of the more general scaling condition: $2 \alpha_F + \alpha_G=0$.} 
for the power exponent, $\alpha_F$, in (i).

If $\alpha_F>0$ is assumed, one would be also left with the contradictory scaling 
condition that $\alpha_F=-1$\cite{Boucaud:2005ce,Boucaud:2008ky} and should conclude that no 
solution exists for that case.

\subSection{Regular case: decoupling solution}

When $\alpha_F=0$, the previous ressumation cannot be done and 
the integral $I(k^2)$ in \eq{LamInf} should be consistently expanded  
in powers of $k^2$. 
This case has been deeply studied in ref~\cite{Boucaud:2010gr}, where 
it is found that 

\beq
I_{\rm IR}(k^2) 
\ \simeq \ 
\delta^2 \ \frac{A(\mu^2) B(\mu^2)}{64\pi^2} \ 
\frac{k^2}{M^2} \ 
\left[ \ln{\frac{k^2}{M^2}} - \frac 5 6 
\ + \ {\cal O}\left(\frac{M^2}{q_0^2}\right) 
\right] \ + \ 
{\cal O}\left(\frac{k^4}{M^4},\delta^4\right) \ .
%\nonumber \\
\label{demoF}
\eeq

\noindent
Then, the ghost dressing  function, including its first correction to the 
leading constant term, should behave as~\footnote{It should be also noted 
that \eq{demoF} imples to take $M^2/q_0^2 \ll 1$. However, 
any correction to that approximation will not play at the order of 
the coefficient eqs.~(\ref{coefC2}), that will keep the same 
value disregarding that of $M^2/q_0^2$, but at the order of the gluon 
mass, $M^2$, inside the logarithm, presummably like the UV part of the 
integral $I(k^2)$ that should be proportional to $k^2$ and 
vanish at least like $1/\log{(q_0^2/\Lambda^2_{\rm QCD})}$) 
when $q_0^2/\Lambda^2_{\rm QCD} \gg 1$.}

\beq\label{solFIRJo}
F_R(q^2) = F_R(0) \left( 1 \ + \ 
\frac{N_C H_1}{16 \pi} \ \overline{\alpha}_T(0) \ 
\frac{q^2}{M^2} \left[ \ln{\frac{q^2}{M^2}} - \frac {11} 6 \right]
\ + \ {\cal O}\left(\frac{q^4}{M^4} \right) \right)
\eeq

\noindent where

\beq\label{coefC2}
\overline{\alpha}_T(0) = \lim_{q \to 0} \left(q^2 + M^2 \right) \frac{\alpha_T(q^2)}{q^2} 
\ = \ M^2 \frac{g^2_R(\mu^2)}{4 \pi} 
F_R^2(0) \Delta_R(0) ,
\eeq
such that the \eq{SDRS} could be satisfied. It should again understood that 
the subtraction momentum for all the renormalization quantities is $\mu^2$. 
In \eq{coefC2}, $\alpha_T=g_T^2/(4\pi)$ is the perturbative strong coupling 
defined in this Taylor scheme~\cite{Boucaud:2008gn}, while $\overline{\alpha}_T$ 
is the extension of the non-perturbative effective charge definition from 
the gluon propagator~\cite{Aguilar:2008fh} to the Taylor 
ghost-gluon coupling~\cite{Aguilar:2009nf}. 
As a consequence of the appropriate {\it amputation} of a 
massive gluon propagator, where the gluon mass scale is the same RI-invariant 
mass scale appearing in \eq{gluonprop}, 
this Taylor effective charge is frozen at low-momentum and gives 
a non-vanishing zero-momentum value in terms of which the ghost-dressing-function 
subleading correction can be expressed.

\Section{Comparison with numerical results from coupled PT-BFM DSE's}
\label{comparing}
%\alinea

We shall now compare the formulas given by eqs.~(\ref{gluonprop},\ref{solFIRJo}) 
with some numerical results for the gluon propagator 
and ghost dressing function. The aim of the comparison is twofold: 
testing the asympotical solution we obtained in the previous section, but 
also checking the consistency of a massive-gluon solution and determining 
the gluon mass as the best-fit parameter in the comparison.
In particular, we will consider the solutions of the coupled system
of gluon and ghost DS equations obtained by applying the pinching technique 
in the background field method (PT-BFM)~\cite{Binosi:2002ft} 
(see also \cite{Binosi:2009qm} and references therein) 
to compare with. This PT-BFM framework leaves us with an attractive model for gluon 
and ghost propagators providing quantitative description of lattice 
data~\cite{Aguilar:2008xm,Aguilar:2010gm} and giving well account of their main qualitative features: 
finite gluon propagator and finite ghost dressing function at zero-momentum. 
Futhermore, the coupled DSE system can be solved with different boundary 
conditions (see below), the solutions compared with the analytical formula and 
how their behaviour depends on these boundary conditions can be thus 
properly studied. This last is the main purpose of this note.

\begin{figure}[hbt!]
\begin{center}
\begin{tabular}{cc}
\includegraphics[width=7cm]{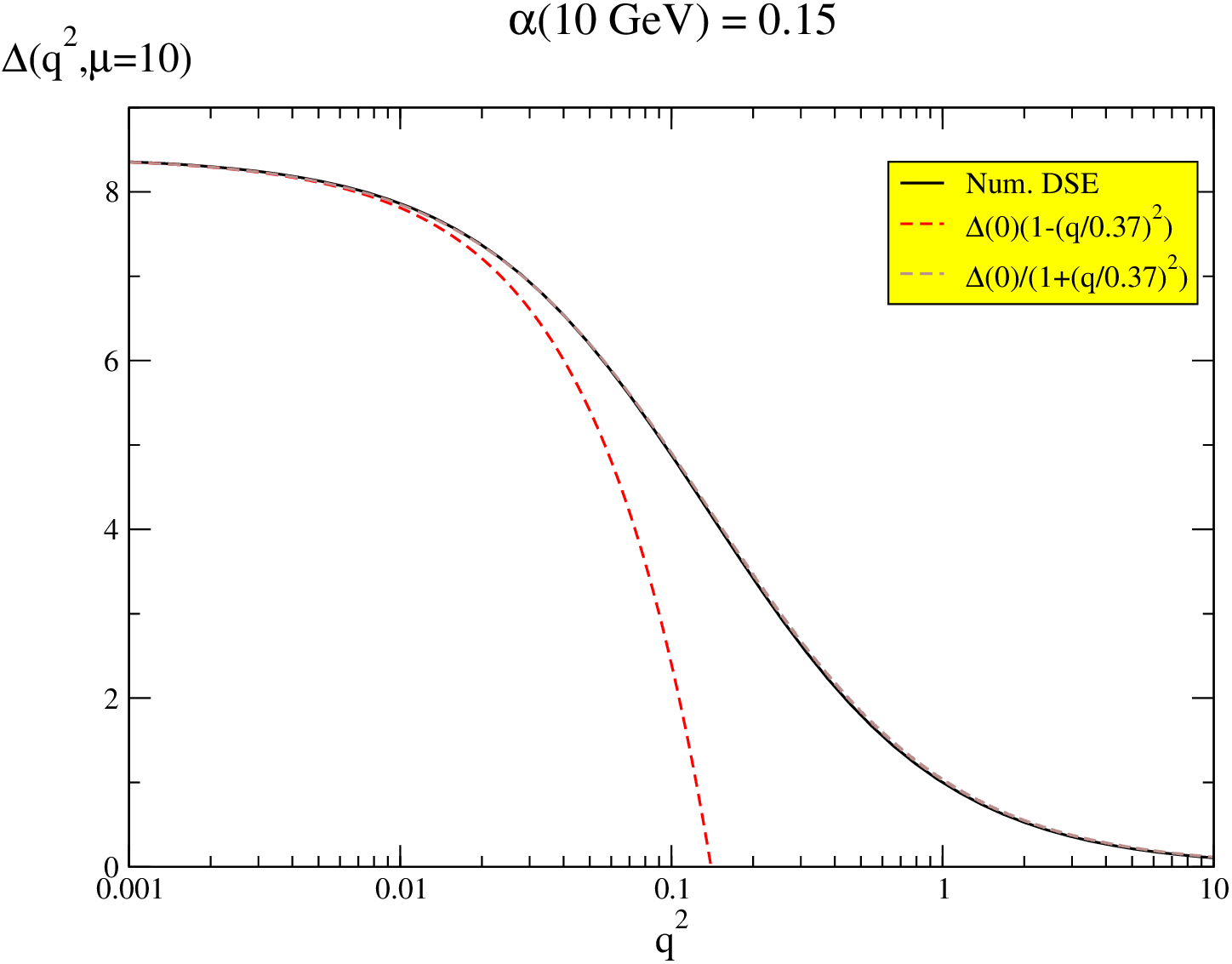} &
\includegraphics[width=7cm]{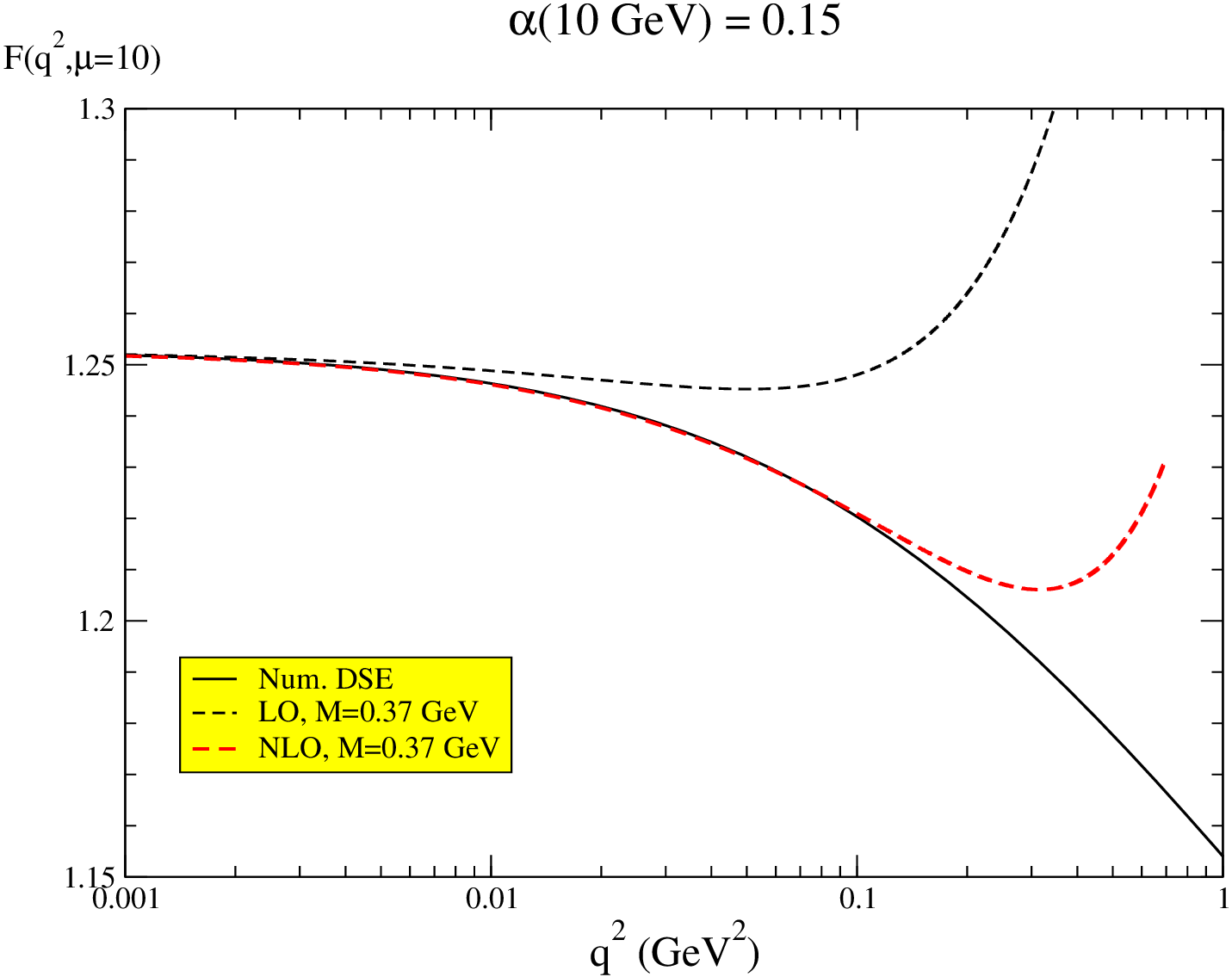} \\
\includegraphics[width=7cm]{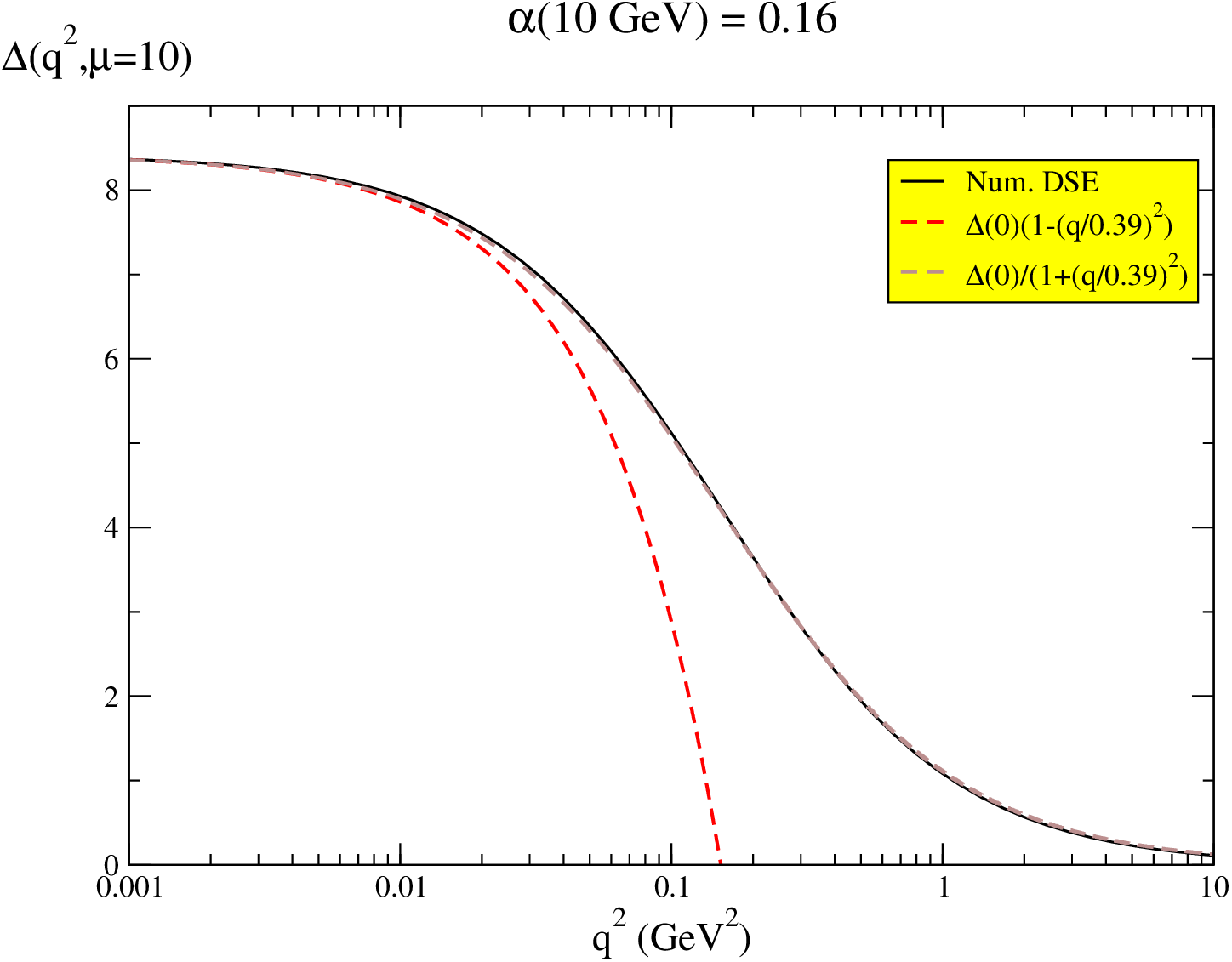} &
\includegraphics[width=7cm]{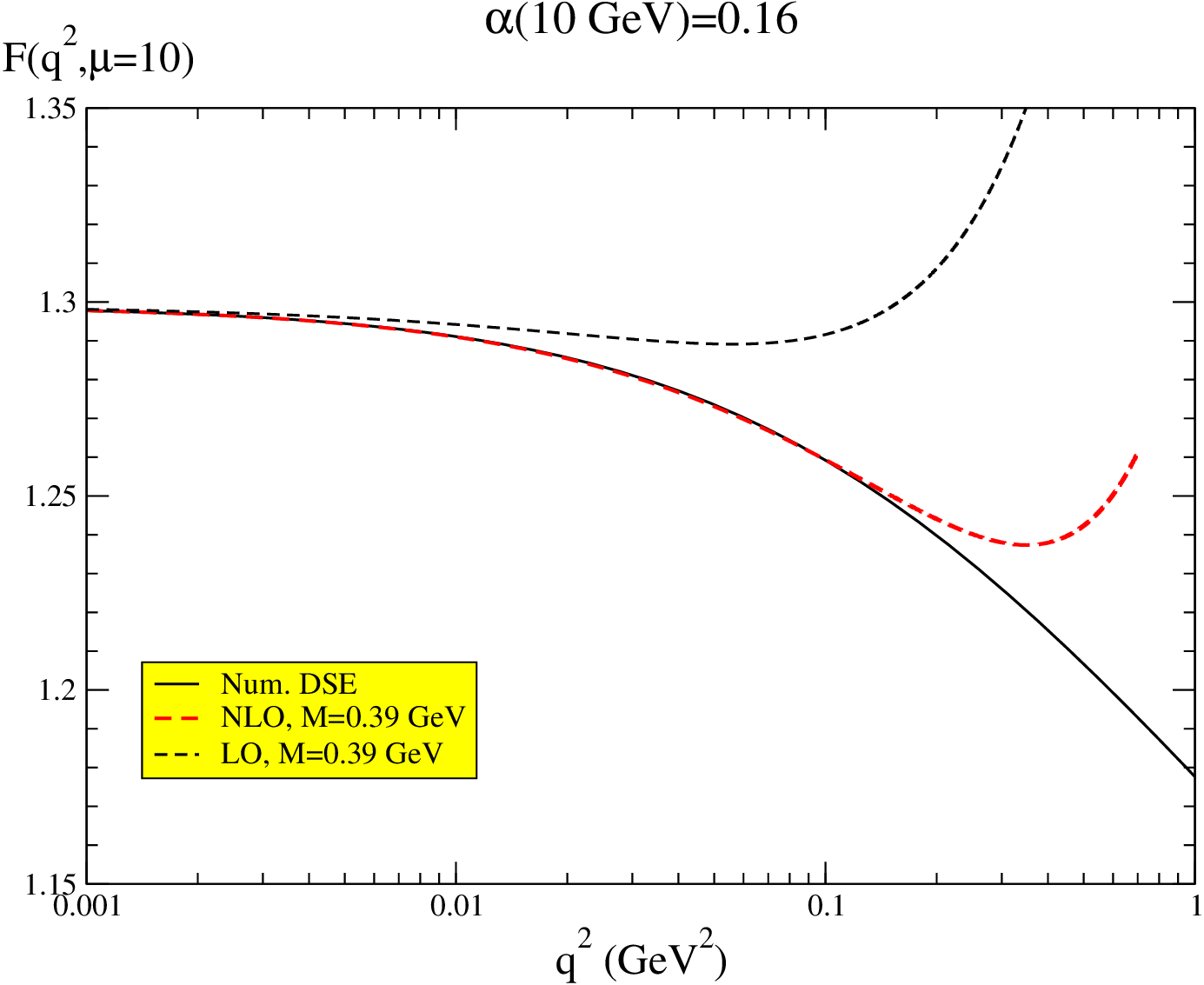} \\
\includegraphics[width=7cm]{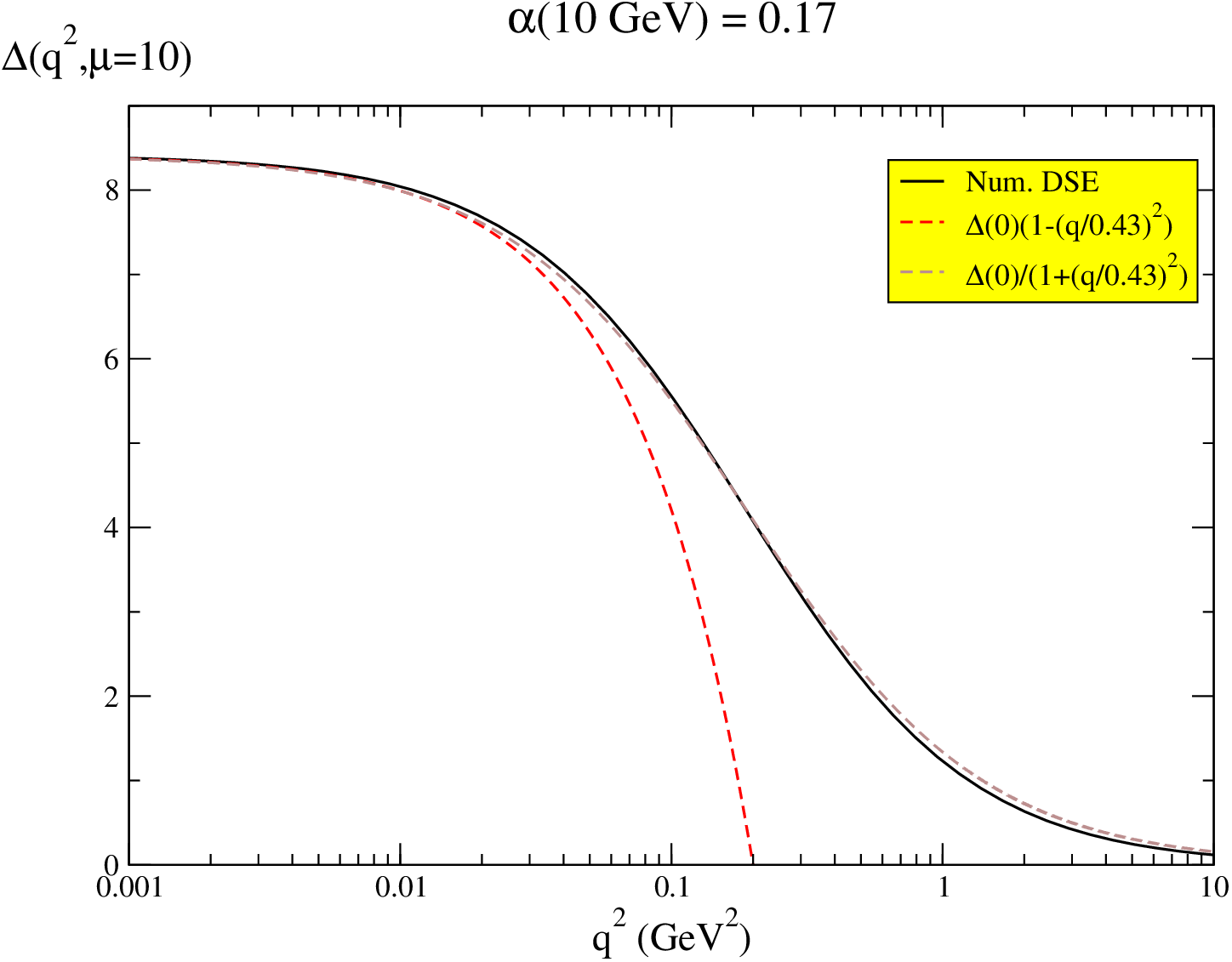} &
\includegraphics[width=7cm]{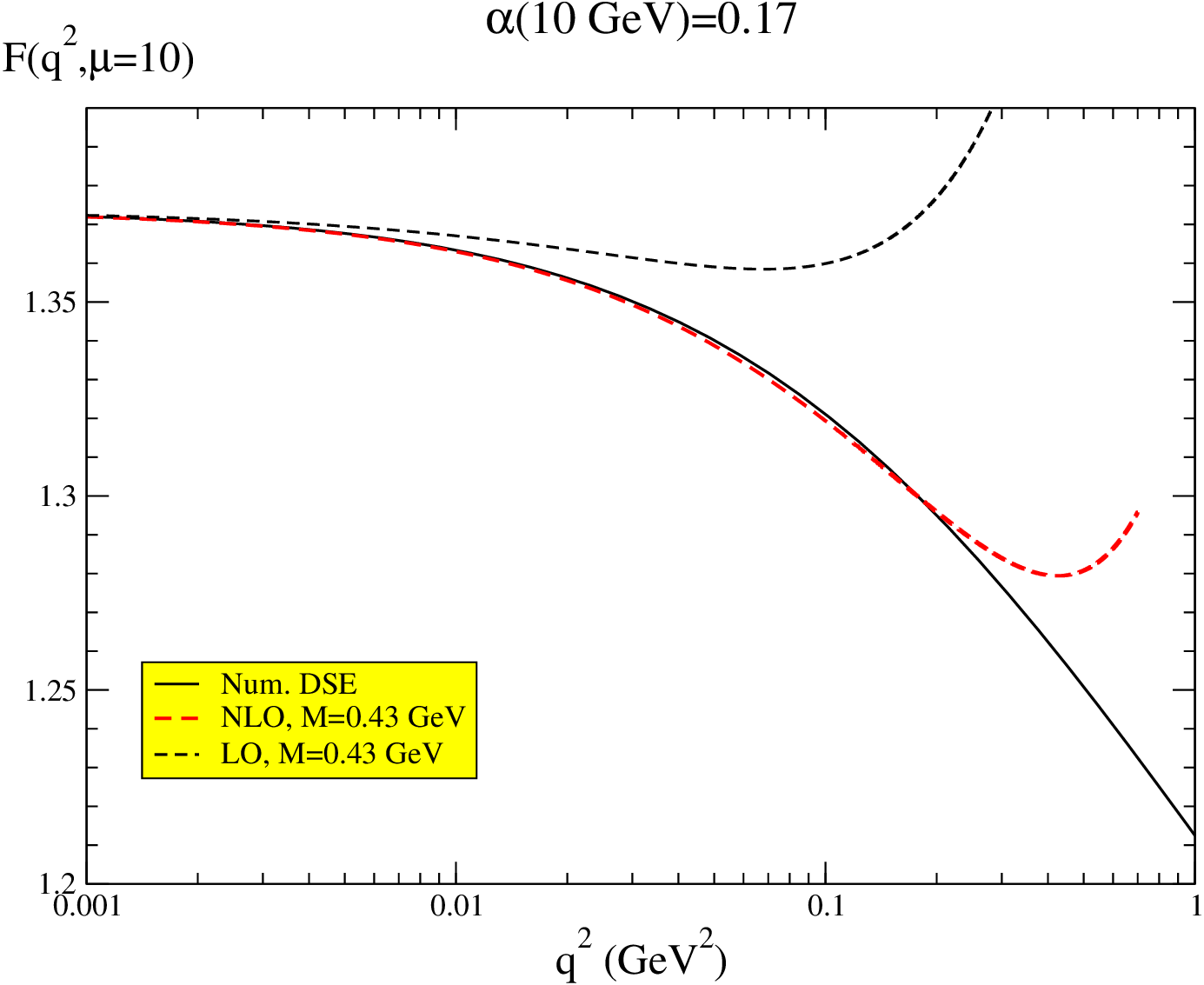} \\
\end{tabular}
\end{center}
\caption{\small Gluon propagators (left) and ghost dressing functions (right) after the numerical integration of  
the coupled DSE system for $\alpha(\mu=10 \mbox{\rm GeV})=0.15,0.16,0.17$ taken from \cite{Aguilar} . The curves 
for the best fit of \eq{gluonprop} to the gluon propagator data appear as dotted lines in the lefthand plots. 
In the righthand plots, the red dotted lines correspond to apply \eq{solFIRJo} with the gluon mass obtained 
from the gluon fits and with $R=\overline{\alpha}_T(0)/M^2$ determined by the zero-momentum values of 
gluon propagator and ghost dressing function coming from the numerical integration of the DSE system; the 
same for the black dotted lines but retaining only the logarithmic leading term in \eq{solFIRJo} 
by dropping the $-11/6$ away.}
\label{fig:ghgl}
\end{figure}

The main feature in the PT-BFM scheme is that the transversality of the gluon self-energy 
is guaranteed order-by-order in the dressed-loop expansion, this leading 
to a gauge-invariant truncation of the gluon DSE~\cite{Binosi:2002ft}.
In this PT-BFM scheme for the coupled DSE system, the ghost propagator DSE 
is the same as given by eqs.~(\ref{ghostSD},\ref{SDRS}), where the bare ghost-gluon 
vertex is approximated by $H_1=1$. The gluon DSE is given by 
\beq\label{coupledDSE}
\frac{(1+G(q^2))^2}{\Delta(q^2)} \left( g_{\mu\nu} - \frac{q_\mu q_\nu}{q^2} \right) =   
q^2 g_{\mu\nu} - q_\mu q_\nu + i \sum_{i=1}^4 \left( a_i \right)_{\mu\nu}
\eeq
where
\beq\label{gluondiags}
a_1 = \gluonSDi ,  & a_2 = \gluonSDii 
\nonumber \\
a_3 = \rule[0cm]{0cm}{1.5cm} \gluonSDiii , & a_4= \gluonSDiv .
\eeq
In the diagrams of (\ref{gluondiags}) for the gluon DSE, \eq{coupledDSE}, the external gluons 
are treated, from the point of view of Feynman rules, as background fields 
(these diagrams should be also properly regularized, as explained in \cite{Binosi:2009qm}). 
The last justifies the four field coupling of two background gluons and two ghosts leading to 
the contribution $a_4$. The function $1+G$ defined in ref.~\cite{Grassi:1999tp} can be, 
in virtue of the ghost propagator DSE, connected to the ghost propagator~\cite{Aguilar:2009nf}. 
The coupled system is to be solved, by numerical integration, 
with the two following boundary conditions as the only required inputs:
the zero-momentum value of the gluon propagator and that of the coupling at 
a given perturbative momentum, $\mu=10$ GeV, that will be used as 
the renormalization point. The latter can be done by fixing
different values for the boundary conditions, this providing us 
with a family of gluon and ghost propagators solutions so determined.
In particular, solutions obtained by keeping the zero-momentum value of the gluon propagator fixed 
(see lefthand plots of fig.~\ref{fig:ghgl}) while $\alpha(\mu^2=100\mbox{\rm~GeV}^2)$ is ranging 
from 0.15 to 0.1817 are available~\cite{Aguilar} and can be confronted to the asymptotical 
expressions derived in the previous section.

\subSection{Decoupling solutions in the PT-BFM scheme}

Then, as the gluon propagator solutions in the PT-BFM scheme result to behave as massive ones, 
the eqs.~(\ref{gluonprop},\ref{solFIRJo}) must account for the low-momentum behaviour of 
both gluon propagator and ghost dressing function with $H_1=1$ and 
\beq\label{alphaT033}
\overline{\alpha}_T(0) = \alpha_T(\mu^2) F_R^2(0) B(\mu^2) \ = \
\alpha_T(\mu^2) F_R^2(0) M^2 \Delta_R(0) \ ,
\eeq
$\alpha_T(\mu^2)=g_R^2(\mu^2)/(4\pi)$ being fixed, as a boundary condition, at the moment 
of the numerical integration of the coupled DSE for each particular solution of the family. 
$B(\mu^2)$ and $M$ being determined by the best fit of the \eq{gluonprop} to the numerical solution 
for the gluon propagator, we shall be left with one only free parameter, 
$F_R(0)$, to account with \eq{solFIRJo} for the numerical solution 
for the ghost propagator. 
Furthermore, the zero-momentum values of the ghost dressing function, $F_R(0)$, can be also 
taken from the numerical integration of the DSE (for any value of the 
$\alpha(\mu=10 \mbox{\rm GeV})$); and these altoghether with the zero-momentum values of the 
gluon propagator, $\Delta_R(0)$, and the gluon masses, obtained by the fit of \eq{gluonprop} to the 
numerical DSE gluon propatator solutions, provide us with all the ingredients 
to evaluate, with no unknown parameter, \eq{solFIRJo}. 
The gluon masses obtained from the best fits of \eq{gluonprop} to the numerical data (see 
the left plots in fig.~\ref{fig:ghgl}) and that of zero-momentum Taylor effective charge, 
$\overline{\alpha}_T(0)$, computed by applying \eq{alphaT033}, with 
the zero-momentum ghost dressing function taken from numerical data, 
can be found in tab. \ref{tab-fits}.

%%%%%%%%%%%%%%%%%%%%%%%%%%
%%%%%%%%%%%%%%%%%%%%%%%%%%

Indeed, the expression given by \eq{solFIRJo} can be succesfully
applied to describe the solutions all over the range of coupling values, 
$\alpha(\mu)$, at $\mu=10$ GeV. This can be seen, for instance, for $\alpha=0.15,0.16,0.17$, 
in the right plots of fig.~\ref{fig:ghgl}, where the ghost dressing functions 
obtained from the numerical integration of the DSE's appear plotted with black-solid 
lines and the evaluation of \eq{solFIRJo}, as we explained above, with red-dotted ones.
Thus, {\it with no parameter to be fitted ($F_R(0)$ is taken from the numerical integration), 
we nicely reproduce the low-momentum behaviour of the ghost dressing function obtained 
through numerical integration}.

In the right plots of fig.~\ref{fig:ghgl}, black-dotted curves obtained by only retaining 
the logarithmic leading order in \eq{solFIRJo} appear also drawn. The big discrepancy they 
show with respect to the numerical integration clearly implies the necessity of the 
next-to-leading correction in \eq{solFIRJo} when assessing the gluon mass from the 
low-momentum ghost propagator.

%%%%%%%%%%%%%%%%%%%%%%%%%%
\begin{table}
\begin{center}
\begin{tabular}{|c||c|c|} 
\hline
\rule[0cm]{0cm}{0.5cm} $\alpha(\mu)$ & $\overline{\alpha}_T(0)$ & $M$ (GeV) [gluon] \\
\hline
\hline
0.15 & 0.24 & 0.37 
\\
\hline
0.16 & 0.30 & 0.39 \\
\hline
0.17 & 0.41 & 0.43 \\
\hline
\end{tabular}
\end{center}
\caption{\small Gluon masses and the zero-momentum non-perturbative effective 
charges, obtained as explained in the text, which are applied to describe
the gluon propagator and the ghost dressing function numerical data with 
eqs.~(\ref{gluonprop},\ref{solFIRJo}).}
\label{tab-fits}
\end{table}
%%%%%%%%%%%%%%%%%%%%%%%%%%%

\subsection{The ``critical'' limit in the PT-BFM scheme}

There appears to be a {\it critical} value of the coupling, 
$\alpha_{\rm crit}=\alpha(\mu^2)\simeq 0.182$ with $\mu=10\mbox{\rm ~Gev}$, 
above which the coupled DSE system does not converge any longer to a solution~\cite{Aguilar}.
As a matter of the fact, we know from \eq{crit-be} that
the scaling solution implies for the coupling
\beq\label{ap:crit}
g^2_{\rm crit} \ = \ g_R^2(\mu^2) 
\simeq
\frac{10 \pi^2}{3 A^2(\mu^2) B(\mu^2) } \ ,
\eeq
where $B(\mu^2)$ is determined by the gluon propagator solution that is supposed to 
behave as \eq{gluonprop}, and $A(\mu^2)$ by the ghost propagator that should behave 
as
\beq\label{ap:crit2}
F_R(q^2) \ = \ A(\mu^2) \left(\frac{M^2}{q^2}\right)^{1/2} \ ,
\eeq
where again $\mu^2$ is the momentum at the subtraction point.
This is shown in ref.~\cite{Boucaud:2008ji}, where only the ghost propagator 
DSE is solved there after extracting a gluon propagator from the lattice data 
and applying it to build the kernel of the integral, \eq{LamInf}, appearing in 
\eq{SDRS}. In the analysis of ref.~\cite{Boucaud:2008ji}, a ghost dressing 
function solution diverging at vanishing momentum appears to exist and 
verifies eqs.~(\ref{ap:crit},\ref{ap:crit2}), 
while regular or decoupling solutions exist for any $\alpha < \alpha_{\rm crit}$.

Now, we can perform a more complete analysis by studying again the  
dressing function computed by solving \eq{coupledDSE} for  
the different values of the coupling, $\alpha=\alpha(\mu^2)$, at $\mu^2=100$ 
GeV$^2$~\cite{Aguilar}. 
A ghost dressing function at vanishing momentum, $F(0,\mu^2)$, 
diverging as $\alpha \to \alpha_{\rm crit}$ had to be expected, one could 
try the following power behaviour,
\beq
F(0) \ \sim \ (\alpha_{\rm crit} - \alpha(\mu^2))^{- \kappa(\mu^2)} \ ,
\eeq
to describe the vanishing-momentum ghost dressing function in terms of 
the coupling, $\alpha(\mu^2)$. The coefficient $\kappa(\mu^2)$ should be the 
positive critical exponent (depending presummably on the 
renormalization point, $\mu^2$) governing the transition from decoupling 
($\alpha < \alpha_{\rm crit}$) to the scaling ($\alpha = \alpha_{\rm crit}$) 
solutions. 

Our strategy will be to let $\alpha_{\rm crit}$ be a free parameter to be fitted 
by requiring the best linear correlation for $\log[F(0)]$ in terms of 
$\log[\alpha_{\rm crit}-\alpha]$. In doing so, the best correlation 
coefficient is 0.9997 for $\alpha_{\rm crit}=0.1822$, which is pretty close to 
the critical value of the coupling above which the coupled DSE system does not converge 
any more, and then we obtain
\beq
\kappa(\mu^2) = 0.0854(6) \ .
\eeq
%

%%%%%%%%%%%%%%%%%%%%%%%%%
\vspace{0.9cm}
\begin{figure}[!hbt]
\begin{center}
\begin{tabular}{cc}
\includegraphics[width=7.5cm]{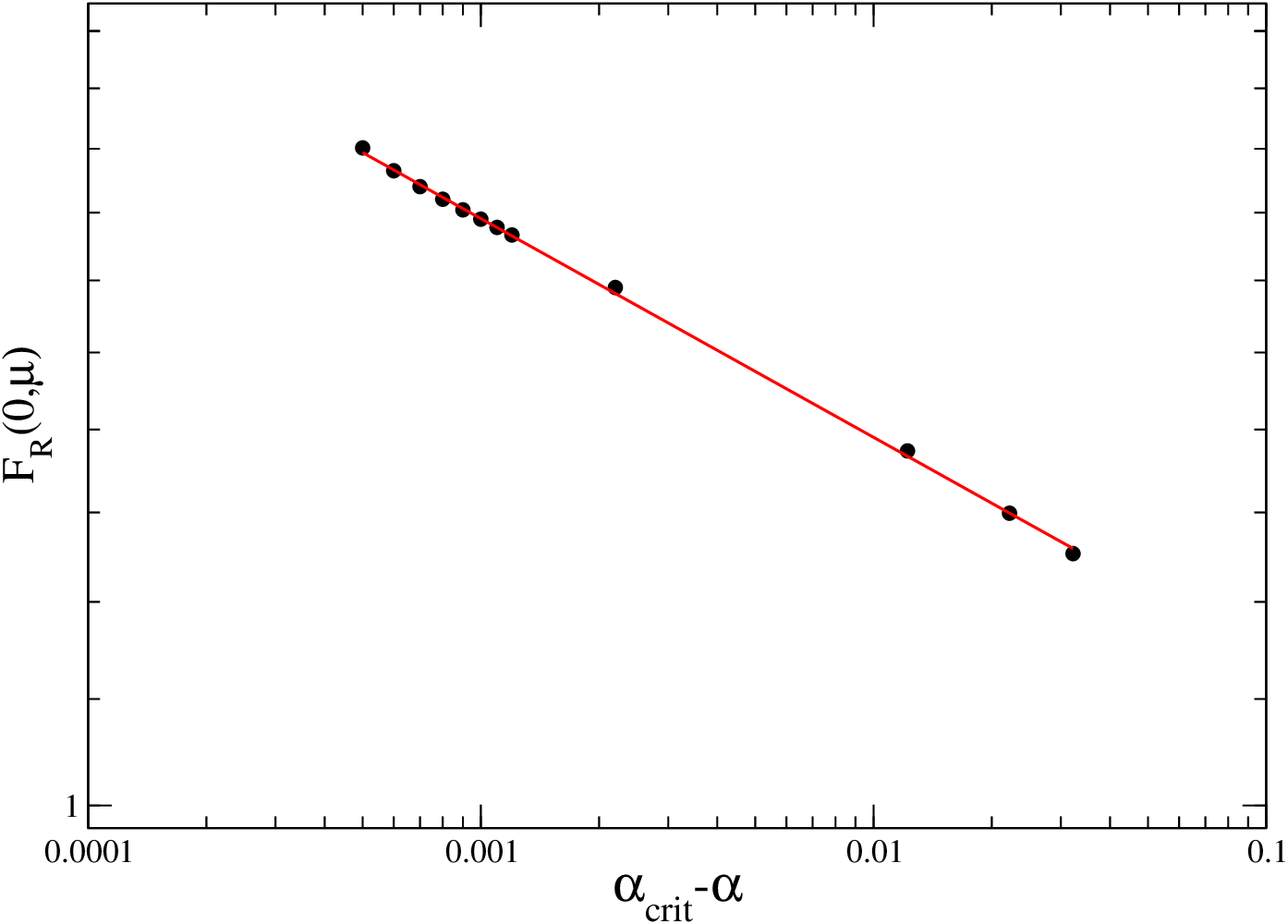} 
&
\includegraphics[width=7.5cm]{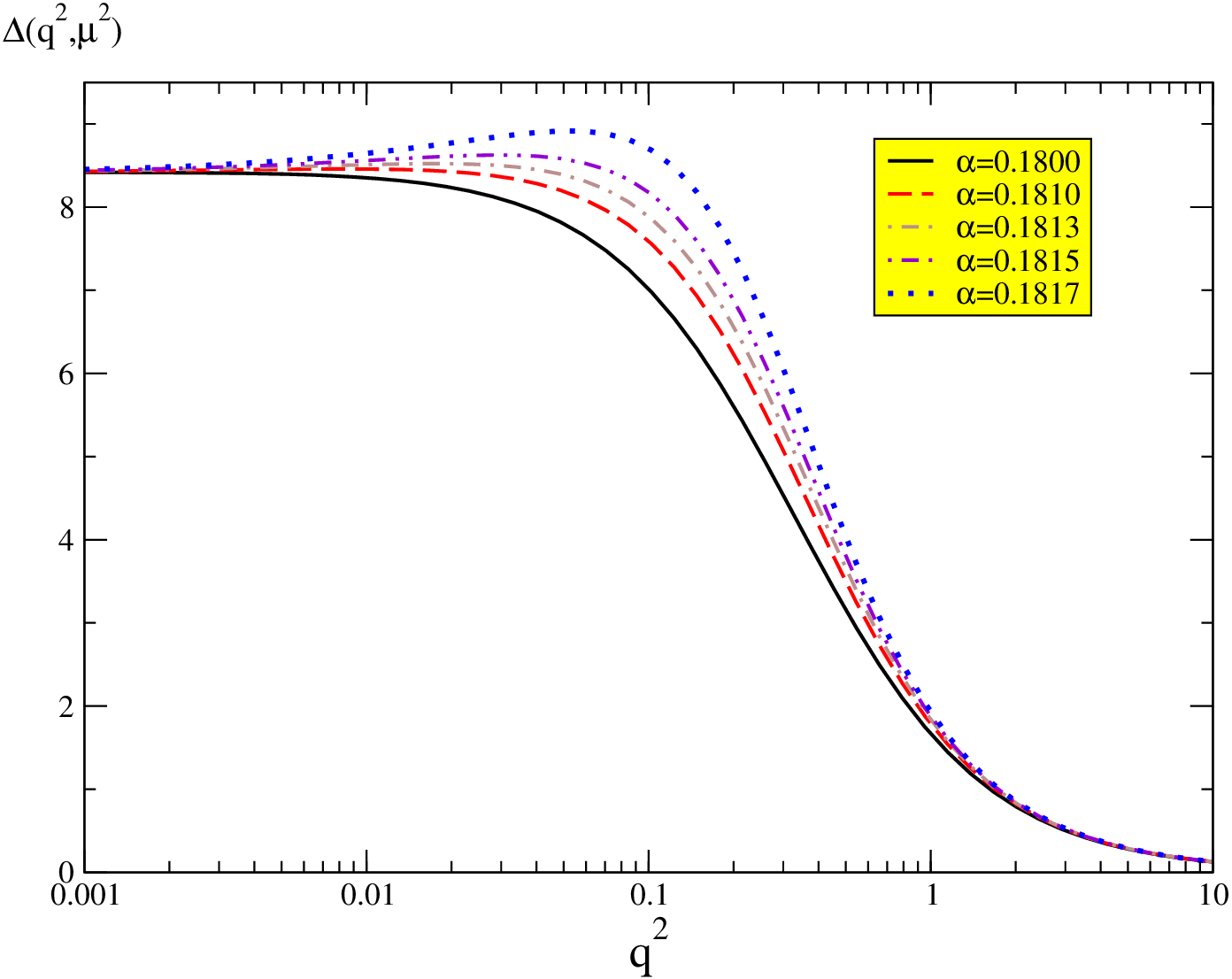} 
\\
(a) & (b) 
\end{tabular}
\end{center}
\caption{\small (a) Log-log plot of the zero-momentum values of the ghost dressing function, obtained 
by the numerical integration of the coupled DSE system in the PT-BFM scheme, in terms of 
$\alpha_{\rm crit}-\alpha$. $\alpha=\alpha(\mu=10 \mbox{\rm GeV})$, 
the value of the coupling at the renormalization momentum, is an initial condition for 
the integration; while $\alpha_{\rm crit}$ is fixed to be 0.1822, as explained in the text, 
by requiring the best linear correlation. %The negative value for the slope indicates that the 
%zero-momentum ghost propagator diverges as $\alpha \to \alpha_{\rm crit}$. 
(b) Gluon propagator solutions
in terms of $q^2$ for the same coupled DSE system for different values of $\alpha(\mu=10 \mbox{\rm GeV})$, 
all very close to the critical value, ranging from 0.18 to 0.1817~.}
\label{fig:ghost0s}
\end{figure}
%%%%%%%%%%%%%%%%%%%%%%%%%

In fig.~\ref{fig:ghost0s}.(a), the log-log plot of $F_R(0)$ in terms of 
$\alpha_{\rm crit}-\alpha$ is shown for $\alpha_{\rm crit}=0.1822$, where 
the linear behaviour corresponding to the best correlation coefficient 
can be strikingly seen and the negative slope indicates a zero-momentum ghost propagator diverging 
as $\alpha \to \alpha_{\rm crit}$. However, {\it no critical or scaling solution of the 
coupled DSE system seems to appear with a massive gluon propagator as solution of the coupled 
DSE system in the PT-BFM, although the decoupling solutions obtained for any 
$\alpha < \alpha_{\rm crit} = 0.1822$ 
seem to approach the behaviour of a scaling one when $\alpha \to \alpha_{\rm crit}$}.  
The absence of the scaling solution can be well understood by analysing \eq{coupledDSE}. 
As explained in \cite{Aguilar:2008xm}, 
after the appropriate regularization and renormalization, the contribution of 
$a_3$ to the inverse of the gluon propagator, its momentum vanishing, will be 
dominated by~\footnote{The regularisation procedure in \cite{Aguilar:2008xm} implies the 
subtraction of the perturbative part, as well as we evaluate \eq{SDRS} for 
two different scales and subtract them in order not to have to deal with any 
UV cut-off in \eq{SDRS}.}
\beq
a_3 \rightarrow \int_0^{q_0} d(q^2) \left( F^2(q^2) - 1 \right) \ ,
\eeq   
where $q_0$ is again some UV cut-off above which the ghost dressing function 
can be taken to be perturbative. Provided that one deals with a decoupling solution,  
the ghost dressing function reaching some constant as $q^2 \to 0$, this contribution 
is finite and negligible (the same happens for $a_4$). On the other hand, had we 
considered the scaling solution, the ghost dressing function would behave as $1/q$ 
and would lead to a divergent contribution and to a vanishing gluon propagator at 
vanishing momentum. Then, a massive gluon propagator, as lattice solutions points to and as 
required in ref.~\cite{Aguilar:2008xm,Aguilar}, cannot appear as a scaling solution. 
The same has been already proven in literature by applying different truncation schemes and also
on general grounds, and indeed agrees with the very well-known infrared behaviour 
obtained from the coupled DSE system in refs.~\cite{Alkofer:2000wg,Fischer:2008uz}, 
where an unique scaling solution with $\alpha_F\simeq - 0.595$ and $\alpha_G=-2 \alpha_F \simeq 1.190$ 
emerges, the gluon propagator vanishing thus at zero-momentum, 
as the inverse of the zero-momentum ghost dressing is assumed to vanish.
However, for the sake of completeness, a very general argumentation addressing this 
issue is presented in appendix \ref{C}. 
We should recall at this point that the numerical analysis of the ghost propagator 
DSE in ref.~\cite{Boucaud:2008ji} left us with a divergent ghost dressing function for 
the critical value of the coupling, even after assuming a finite gluon propagator at zero 
momentum. However, this resulted from a partial analysis where we dealt only with 
the ghost propagator DSE and not with the gluon propagator one. 
Thus, we did not solve the gluon DSE and taking a massive gluon propagator from the 
lattice to build the ghost-DSE kernel does not 
prevent from obtaining a ``{\it wrong}'' scaling solution indeed not satisfying the coupled 
DSE system.

When approaching the critical value of the coupling, the gluon propagators obtained 
from the coupled DSE system in PT-BFM must be also thought to obey the 
same critical behaviour pattern as the ghost propagator. 
In the PT-BFM, the value at zero-momentum being 
fixed by construction~\cite{Aguilar:2008xm,Aguilar}, one should expect that, 
instead of decreasing, the 
gluon propagator obtained for couplings near to the critical value increases for low momenta: 
the more one approaches the critical coupling the more it has to increase. 
This is indeed the case, as can be seen in fig.~\ref{fig:ghost0s}(b). 
This implies that, near the critical value, the low momentum propagator does not 
obey \eq{gluonprop} and that consequently \eq{solFIRJo} does not work any longer to 
describe the low momentum ghost propagator\footnote{This is only true for the 
next-to-leading contribution of \eq{solFIRJo}, the leading one being 
only determined by the zero-momentum gluon propagator still works.}.

Finally, one can pay attention to the critical value of the coupling, $\alpha_{\rm crit}=0.1822$, 
and try to make a comparison with the physical strong coupling values in order get some idea of 
whether the current data can exclude or not this critical behaviour. Although the experimental PDG {\it world 
average} of the strong coupling in the $\overline{\rm MS}$ scheme, $\alpha_{\overline{\rm MS}}(M_Z)=0.1184(7)$~\cite{PDG}, 
can be propagated from the $Z^0$ boson mass down to $\mu=10$ GeV 
to give $\alpha_{\overline{\rm MS}}(10 \ {\rm GeV})=0.179(2)$, that 
incidentally lies on the right ballpark of the above critical value, such a comparison 
is meaningless because our coupling corresponds to one in MOM Taylor-scheme for 
zero number of flavours. One can use instead the available perturbative four-loop formula describing 
the running of the coupling in Taylor-scheme to estimate $\Lambda_{\rm QCD}$ in this particular scheme, 
then perform the conversion to $\overline{\rm MS}$ (see for instance eqs.(22,23) 
of the first reference in~\cite{Boucaud:2008gn}) and thus obtain the value quoted in tab.~\ref{tabLambda}. 
Of course, it would be again meaningless to compare this last value with the one 
for $\Lambda_{\overline{\rm MS}}$ that can be 
obtained from the PDG value for $\alpha_{\overline{\rm MS}}(M_Z)$, also quoted in tab.~\ref{tabLambda}, 
but we can refer the comparison to the lattice Yang-Mills determinations of the same parameter\footnote{It should 
be noted that the procedures for the lattice determination of $\Lambda_{\overline{\rm MS}}$ mainly work in 
the UV domain, where IR sources of uncertainties as the Gribov ambiguity or volume effects are indeed negligible. 
In fact, there are unquenched lattice determinations with $N_f=5$ staggered fermions for the strong 
coupling~\cite{Davies:2008sw} which 
are pretty consistent with the PDG value.}, as for instance the two of them included in tab.~\ref{tabLambda}. 
Thus, the lattice estimates of 
$\Lambda_{\overline{\rm MS}}$ appear to lie clearly below this critical limit for the PT-BFM DSE  
in pure Yang-Mills. However, as no quark flavour loops effect have been incorporated in our   
DSE, \eq{coupledDSE}, we cannot neither compare with the physical strong coupling nor conclude 
whether the critical limit can be allowed in the ``{\it real world}''.

%------------------------------------------------
\begin{table}[htb!]
\begin{center}
\begin{tabular}{|c|c|c|c|}
\hline 
$\Lambda_{\overline{\rm MS},{\rm crit}}^{N_f=0}$ 
& 
$\Lambda_{\overline{\rm MS}}^{N_f=0}$~\cite{Luscher:1993gh} 
& 
$\Lambda_{\overline{\rm MS}}^{N_f=0}$~\cite{Boucaud:2008gn}
& 
$\Lambda_{\overline{\rm MS}}^{N_f=5}$~\cite{PDG} 
\\
\hline
434 MeV & 238(19) MeV & 244(8) MeV & 213(9) MeV \\
\hline
\end{tabular}
\end{center}
\label{tabLambda}
\caption{\small The critical value of $\Lambda_{\overline{\rm MS}}$ in pure Yang-Mills inferred 
from $\alpha_{\rm crit}=0.1822$ (first column), lattice estimates for Yang-Mills $\Lambda_{\overline{\rm MS}}$ 
taken from literature (second and third columns) and the one obtained from the PDG value 
of $\alpha_{\overline{\rm MS}}(M_Z)$ by applying a four-loop perturbative formula for the running of  
$\alpha_{\overline{\rm MS}}$ with $N_f=5$.} 
\end{table}
%-------------------------------------------------

\vspace*{1cm}

\Section{Conclusions}\label{conclu}
%\alinea

The ghost propagator DSE, with the only assumption of taking $H_1(q,k)$ from the 
ghost-gluon vertex in \eq{DefH12} to be constant in the infrared domain of $q$, can be 
exploited to look into the low-momentum behaviour of the ghost propagator.  The
two classes of solutions named ``decoupling'' and ``scaling'' can be indentified and 
shown to depend on whether the ghost dressing function achieves a finite non-zero
constant ($\alpha_F=0$) at vanishing momentum or not ($\alpha_F \neq 0$). The 
solutions appear to be dialed by the size of the coupling at the renormalization 
momentum which plays the role of a boundary condition for the DSE integration. 
The low-momentum behaviour of the decoupling solutions results to be regulated by the 
gluon propagator mass and by a regularization-independent 
dimensionless quantity that appears to be the effective charge defined from 
the Taylor-scheme ghost-gluon vertex at zero momentum.

In this note, we have studied the solutions of coupled ghost and gluon propagator DSE in 
the PT-BFM scheme 
and demonstrated that the asymptotic decoupling formula ($\alpha_F=0$) 
successfully describes the low-momentum ghost propagator. 
The model applied for the massive gluon propagator is also verified to give properly account 
of the gluon solution, at least for momenta below 1 GeV (and for a coupling 
not very close to the critical point).
Although we argued that a massive gluon propagator implies that the ghost dressing 
function takes a non-zero finite value at vanishing momentum, 
we also show that the zero-momentum ghost dressing function 
tends to diverge when the value of the coupling dialing the solutions 
approaches some critical value. Such a divergent behaviour at the critical coupling 
seems to be the expected one for a scaling solution (where, if the gluon 
is massive, $\alpha_F=-1/2$). If we consider the zero-momentum value of the ghost 
dressing function as some sort of ``{\it order parameter}'' indicating whether the 
ghost propagator low-momentum behaviour is suppressed ($\alpha_F=0$ and finite ghost 
dressing function) or it is enhanced ($\alpha_F<0$ and divergent ghost dressing 
function), the strength of the coupling computed at some renormalization point 
seems to control some sort of transition from the {\it suppressed} to 
the {\it enhanced} phases for the ghost propagator DSE solutions in the PT-BFM scheme.
The last only takes place as some critical value of the coupling is reached. 
Neverteless, it can be proven that, as far as the gluon is massive,  
the scaling behaviour for the Yang-Mills propagators appear not to be a 
solution but an unattainable limiting case for the PT-BFM DSE solutions. 

\bigskip

{\bf Acknowledgements:} 
The author is particularly indebted to Ph.~Boucaud, J.P~Leroy, A.~Le~Yaouanc, J.~Micheli and 
O.~P\`ene for very fruitful discussions at the initial stages of the work and to 
J.~Papavassiliou and A.C.~Aguilar also for very valuable discussions and comments, and 
specially for providing me with some unpublished results 
which were exploited in this paper. J. R-Q also acknowledges the Spanish MICINN for the 
support by the research project FPA2009-10773 and ``Junta de Andalucia'' by P07FQM02962.

%\newpage

\appendix

\Section{No scaling solution with massive gluons}
\label{C}

We consider the conventional gluon self-energy, $\Pi^{\mu\nu}(q)$, 
contributing to the gluon DSE:

\beq\label{ap:cDSEs}
\Pi_{\mu\nu}(q)
&=&
\rule[0cm]{0cm}{1cm}
\frac 1 2 \gluonSDi%
+ \ 
\frac 1 2 \gluonSDiib%
+  
\gluonSDiii%
\nonumber \\
&+&
\rule[-0.9cm]{0cm}{2.1cm}
\frac 1 6 \gluonSDv%
+ \ 
\frac 1 2 \gluonSDvi ,
\eeq
where the yellow bullets stand for full vertices and propagators. 

After only assuming that the ghost-gluon 
vertex form factor $H_1$ is constant (the Taylor non-renormalization 
theorem tells us that the bare ghost-gluon vertex is finite), 
We showed in section \ref{revisiting} that a massive 
gluon propagator unequivocally implies 
a ghost dressing function diverging as $1/q$ at vanishing momentum.
Then, the ghost-loop contribution to the gluon self-energy, \eq{ap:cDSEs}, 
with vanishing external momentum, $k$, is dominated by
\beq\label{ap:div}
g^{T}_{\mu\nu}(k)  \times \rule[0cm]{0cm}{1cm} \gluonSDiii  
&\sim& 
\int \frac{d^4q}{(2\pi)^4} 
 q^2 \left(1 - \frac{(k\cdot q)^2}{k^2 q^2} \right) 
\frac{F(q^2)}{q^2} \frac{F((q-k)^2)}{(q-k)^2}
\nonumber \\
&\sim&
\int_0^{q_0} dq \ \frac{q^2}{\left(q^2+k^2\right)^{3/2}} 
\int_0^\pi d\theta  \ \frac{\sin^4{\theta}}{\left(1-\frac{2kq}{q^2+k^2} \cos{\theta}\right)^{3/2}}
\nonumber \\
&\sim&
\int_0^{q_0} dq \ \frac{q^2}{\left(q^2+k^2\right)^{3/2}} 
%\sim \ln{\left(\frac{q_0+\sqrt{k^2+q_0^2}}{k}\right) } \;
\eeq
where $q_0$ is the momentum scale which the ghost dressing function is assumed 
to take its asymptotic infrared form below. For not to have to deal again 
with any UV regularization cut-off, we can consider the two momenta $p,k$ such 
that $k^2 \ll p^2 < q_0^2$ and subtract the gluon DSE for these two momenta.
Then, when $k^2 \to 0$ while $p^2$ is kept fixed, one would have for the 
transversal gluon propagator
\beq
\frac 1 {\Delta_R(k^2)} - \frac 1 {\Delta_R(p^2)} \ \sim \ 
\int_0^{q_0} dq \ \frac{q^2}{\left(q^2+k^2\right)^{3/2}} \ \ ,
\eeq
where we only account for the dominant part of the ghost-loop contribution.  
This contribution diverges thus logarithmically with 
a constant ghost-gluon vertex again as the only assumption made, while
computing the contributions coming from other diagrams 
would leave us with the neccessity to make some new assumption about the 
full gluon vertices. 

Thus, to avoid a diverging gluon self-energy, we need to invoke 
new contributions from the other diagrams in \eq{ap:cDSEs}, 
also diverging logarithmically as 
the external momentum vanishes, to cancel that from \eq{ap:div}. 
Otherwise, such a divergent behaviour of the gluon self-energy would 
lead the inverse of the gluon propagator to diverge, and  
the gluon propagator consequently to vanish, at zero-momentum.

%%%%%%%%%%%%%%%%%%%%%%%%%%%%%%%%%%%%%%%%%%%%%%%%%%%%%%%%%
%%                              BIBLIO
%%%%%%%%%%%%%%%%%%%%%%%%%%%%%%%%%%%%%%%%%%%%%%%%%%%%%%%%%

%%%%%%%%%%%%%%%%%%%%%%%%%%%%%%%%%%%%%%%%%%%%%%%%%%%%%%%%%

%%%%%%%%%%%%%%%%%%%%%%%%%%%%%%%%%%%%%%%%%%%%%%%%%%%%%%%%%%%%%%%%%%%%
%%%%%%%%%%%%%%%%%%%%%%%%%%%%%%%%%%%%%%%%%%%%%%%%%%%%%%%%%%%%%%%%%%%%
%
%
\end{document}